\documentstyle[pra,preprint,aps]{revtex}

\tighten
\draft

\begin{document}

\title{
Quantifying entanglement\\ 
in\\
 ``Experimental entanglement of four particles'' }
\author{ {\bf Arun K. Pati} $^{(1)}$ }
\address{Institute of Physics, Sainik School Post, Bhubaneswar-751005,
Orissa, India}
\address{$^{(1)}$ School of Informatics, University of 
Wales, Bangor LL 57 1UT, UK}
\address{Email: akpati@iopb.res.in}


\maketitle
\def\ra{\rangle}
\def\la{\langle}
\def\ver{\arrowvert}

\begin{abstract}
We quantify various possible entanglement measures for the
four-particles GHZ entangled state that has been produced experimentally
[C. Sackett {\it et al}, Nature {\bf 404}, 256-259 (2000)].
\end{abstract}

\vskip 1cm

In quantum world entangled states naturally arise due to tensor 
product structure of the composite Hilbert space and the linear 
superposition principle. This was first realized by Schr{\"o}dinger
as the quintessential feature of quantum mechanics that has no classical
analog. With the recent advances in quantum computation and information theory
quantum entanglement is in the lime light as it can do real wonders.
Entangled
states have been regarded as `resources' in quantum information processing,
without which many important tasks will be impossible \cite{mike}.

There is a lot of excitement about preparation of multi-particles
entangled states for fundamental reasons and technological applications
\cite{rb}. Following the proposal of Molmer and Sorenson
\cite{ms}, recently a dramatic step in experimental realization of four
ions entangled state has been taken by Sackett {\it et al} \cite{sack}.
Employing laser-cooling and optical pumping techniques with ${^9}Be^+$ ions,
the Sackett group has produced deterministic entanglement, unlike earlier
schemes based on conditional measurement (post-selection) methods which
are probabilistic in nature. In the experimental setup it is aimed to produce
a four-particles GHZ state 
\begin{eqnarray}
\arrowvert \Psi_4 \rangle = \frac{1}{\sqrt 2} (
\arrowvert \uparrow \uparrow \uparrow \uparrow \uparrow \rangle + i
\arrowvert \downarrow  \downarrow \downarrow \downarrow \rangle),
\end{eqnarray}
where $\arrowvert \uparrow \ra = \arrowvert 0 \ra$ and 
$\arrowvert \downarrow \ra = \arrowvert 1 \ra$ are the computational states.
However,
because of the presence of unavoidable decoherence the actual state produced
is a {\em mixed state} of the form 
\begin{eqnarray}
\rho =  0.43 \arrowvert \Psi_4 \rangle \langle \Psi_4 \arrowvert
+ 0.57 \rho_{\rm incoh}, 
\end{eqnarray}
where $\rho_{\rm incoh}$ is the
completely incoherent part \cite{sack}. The state preparation fidelity
in the said experiment 
is $F=  0.57 \pm 0.02$. Therefore, one can write the experimentally prepared
state as a pseudo-pure state
\begin{eqnarray}
\rho =  0.54 \arrowvert \Psi_4 \rangle \langle \Psi_4 \arrowvert
+ 0.46 \frac{I}{16}, 
\end{eqnarray}
where $I$ is the identity matrix for four qubit system.
This guarantees that $F= \langle \Psi_4 \arrowvert \rho \arrowvert
\Psi_4 \rangle =  0.57$. Here, we show
that the above states are indeed non-separable and {\em quantify } 
various possible 
entanglement content of the actual pseudo-pure state produced in the
experiment. To quantify the entanglement of a multi-particle state
is in general a difficult and an open question
\cite{ved1,bprst,hhh}. However, for pseudo-pure states produced in the 
experiment our brief communication
will answer this question.

Since in Sackett {\it et al}'s scheme decoherence is invariably (in fact, the
source of decoherence is not very well understood)  present if one wants
to scale their scheme it is likely that one might end up in producing a
$N$ particle pseudo-pure ionic state. We quantify the amount of entanglement
present for $N$-particle pseudo-pure ionic state when it is not separable.
The $N$-particle pseudo-pure ionic state can be written as
\begin{eqnarray}
\rho = (1 - \epsilon) \frac{I}{2^N} + \epsilon
\arrowvert \Psi_N \rangle \langle \Psi_N \arrowvert,
\end{eqnarray}
with $ \epsilon$ being the purity parameter, $I$ is the identity
operator in $2^N$-dimensional Hilbert space and
\begin{eqnarray}
\arrowvert \Psi_N \rangle = \frac{1}{\sqrt 2} (
\arrowvert \uparrow \uparrow ....\uparrow \rangle + i^{N+1}
\arrowvert \downarrow  \downarrow ....\downarrow \rangle).
\end{eqnarray}
Notice that after tracing out any $(N-1)$ ions from
the pure state component 
the reduced density matrix of the
last ion is $\frac{I}{2}$. It has
degenerate eigenvalues equal to $\frac{1}{2}$.
We can write the $N$-particle pure state component in terms of
Schmidt decomposition  as
$\arrowvert \Psi_N \rangle = \arrowvert \Psi(-) \rangle =
\frac{1}{\sqrt 2} (
\arrowvert {\tilde \uparrow} \rangle \arrowvert \downarrow \rangle
- \arrowvert {\tilde \downarrow}\rangle \arrowvert \uparrow \rangle)$ where 
$\arrowvert {\tilde \uparrow} \rangle, \arrowvert {\tilde \downarrow} 
\rangle$ are orthonormal basis for $(N-1)$ ions and
$\arrowvert \uparrow \rangle, \arrowvert \downarrow \rangle$ are for 
any of the last ion \cite{pb}. 
With this the pseudo-pure ionic state can be expressed as

\begin{eqnarray}
\rho = (1 - \epsilon) \frac{I}{2^N} + \epsilon
\arrowvert \Psi(-) \rangle
\langle \Psi(-) \arrowvert.
\end{eqnarray}

We wish the above density matrix to
live in a $4$-dimensional Hilbert space. 
Let us now project it onto the $4$-dimensional subspace spanned by the set
$\{\arrowvert {\tilde \uparrow}\rangle \arrowvert \uparrow\rangle$,
$\arrowvert {\tilde \uparrow}\rangle \arrowvert \downarrow \rangle$,
$\arrowvert {\tilde \downarrow} \rangle \arrowvert \uparrow \rangle$,
$\arrowvert {\tilde \downarrow} \rangle \arrowvert \downarrow \rangle \}$.
The resulting density matrix is given by

\begin{eqnarray}
{\tilde \rho} =  \frac{2^N}{4 + \epsilon (2^N-4)}
 \bigg[ (1 - \epsilon) \frac{I_4}{2^N}
+ \epsilon
\arrowvert \Psi(-) \rangle
\langle \Psi(-) \arrowvert \bigg]
\end{eqnarray}
which lives in $4$-dimensional Hilbert space.
This state is effectively a Werner state \cite{rfw}.
Now we come to quantifying the amount of entanglement present in a
$N$-particle pseudo-pure GHZ state.
The crucial observation in our method is that
we can transforms any arbitrary
density matrix in $2^N$-dimensional Hilbert space to a one in
$4$-dimensional Hilbert space without
increasing the entanglement content of the state. Once we know that the
state lives in $4$-dimensional Hilbert space we can apply the known 
results to quantify the entanglement of multi-particle state.

It has been argued  \cite{ls} that if a density
matrix lives in $4$-dimensional Hilbert space then $\rho$ has a unique
decomposition given by
$\rho =  \lambda \rho_s + (1 - \lambda) \arrowvert \Psi_e \rangle
\langle \Psi_e \arrowvert$.
The entanglement of such a density matrix is given by

\begin{equation}
E(\rho) =  ( 1 - \lambda) E(\arrowvert \Psi_e \rangle),
\end{equation}
where $ E(\arrowvert \Psi_e \rangle)$ is the von Neumann entropy of the 
entangled pure state component. We can compare our multi-ion density matrix
after local projection as given by (7), which can be expressed as

\begin{eqnarray}
&& {\tilde \rho} = (1 -x) \frac{I_4}{4} + x
\arrowvert \Psi(-) \rangle \langle \Psi(-) \arrowvert \nonumber \\
&& = \lambda \rho_s + (1 - \lambda) \arrowvert \Psi(-) \rangle \langle
\Psi(-) \arrowvert
\end{eqnarray}
where $x = \frac{\epsilon 2^N }{ 4 + \epsilon(2^N- 4)}$, $\lambda =1$
for $x \le \frac{1}{3}$ and $\lambda = \frac{3}{2}(1 -x)$ for $\frac{1}{3}
\le x \le 1$ (for details see \cite{em}).

When we know that the state is not separable
we can define the entanglement measure for $N$-particle pseudo-pure GHZ
state.
To know if $\rho$ is separable or not we can apply
the fidelity criterion \cite{bet,bdsw}. This says that the state 
is separable if $F = \langle \Psi(-) \arrowvert \rho \arrowvert \Psi(-) \rangle
\le \frac{1}{2}$ otherwise non-separable. Using this we can derive a
bound on the purity parameter
above which the density matrix is {\em fully} (irrespective of partitioning)
non-separable.
Since local projection onto a subspace cannot create entanglement,
it follows that if the original unprojected state (6)
is non-separable we must have $\epsilon > {1 \over 1+ 2^{N-1}}$.
Similar bounds have been obtained by other method for arguing the
absence of entanglement in NMR quantum computation \cite{sam}.
In particular, this kind of bound has been also used to prove the
absence of entanglement while implementing Grover's algorithm on 
NMR quantum computers \cite{bp}.
When we have four ions, a pseudo-pure state is non-separable if 
$\epsilon >  0.11111$. Since in the Sackett {\it et al}'s experiment
the value of $\epsilon$ is $0.54$ {\it it is clearly non-separable}.
Though the produced state is not an ideal GHZ state it has some
amount of entanglement. Therefore, it is important to know how much
entanglement could be there in such a state.

From (8) we can say the $N$-particle pseudo-pure ionic state has at least
an amount entanglement is given by

\begin{equation}
E(\rho) =  \frac{ 2^N \epsilon}{ 4 + \epsilon(2^N- 4)}
E(\arrowvert \Psi(-) \rangle),
\end{equation}
where $E(\arrowvert \Psi(-) \rangle)$ is the entanglement
of the pure $N$ particle ionic state. The multiplication factor in the above
expression is due to the error in preparing a pure $N$-particle GHZ state.
Hence,
the amount of entanglement deviates from actual pure state case.
However, one can see that for large $N$, this extra factor tends to unit for
all $\epsilon$. This shows that if one can entangle more and more number of
ions, then even if one is not able to produce an ideal pure $N$-particle
GHZ state, still a pseudo-pure GHZ state will be good enough.

For multi-particle systems there is no unique measure of entanglement (and it
seems very unlikely that there can be one). Since entanglement is a 
`resource' in quantum world, its `measure' will be different depending on 
what we want to do with it and where do we want to spend it. 
If we consider bi-partite partitioning for $N$-particle GHZ state, then there are
$\frac{N}{2}$ possible partitioning. In such a scenario each partitioning has 
a measure of
entanglement $\log 2$. Total entanglement with $N/2$ partitioning will be 
$\frac{N}{2} \log 2$. Therefore, the average entanglement content will be 
$\frac{1}{2} \log 2$ which is independent of $N$. Thus, 
for four ion pseudo-pure GHZ state the amount of entanglement is 
\begin{equation}
 E(\rho) =  \frac{ 8 \epsilon}{ 4 + 12 \epsilon} \log 2.
\end{equation}
Since the purity parameter in the ion trap setup is only $0.54$, this implies
that amount of entanglement present in $4$-particle pseudo-pure ionic 
GHZ state is only $E(\rho) = 0.412 \log 2$.

If we imagine a situation that the $N$-particle GHZ state has been created from
$2$-particle GHZ state by attaching $(N-2)$ particle locally either by Alice or
Bob and performing sequence of CNOT operations, then that will contain 
just $\log 2$ ebits shared between Alice and Bob. This is because this 
$N$-particle GHZ state can be used to teleport just a single qubit. Hence, 
the $N$-particle GHZ state has an entanglement 
$E(\arrowvert \Psi(-) \rangle) = \log 2$. That such a $N$-particle GHZ state
has $\log 2$ (if $\log$ base is $2$ then it is unit) has also been obtained 
via a global measure of entanglement \cite{meyer}.
Then, in this scenario the measure of entanglement $E(\rho)$ will be
\begin{equation}
 E(\rho) =  \frac{ 16 \epsilon}{ 4 + 12 \epsilon} \log 2.
\end{equation}
For the experimental four-ionic state, by this measure, 
$E(\rho) = 0.824 \log 2$.

Let us consider the measure of entanglement of a composite system using 
the operator norm approach \cite{yuka}. This is realized
by the action of an arbitrary operator on disentangled states which generates
entangled states. If $A$ is an arbitrary operator and $A^{\otimes}$ is 
non-entangling operator then $E(A) = \log \frac{ ||A||_{\cal D} }
{ ||A^{\otimes}||_{\cal D} }$, where $E(A)$ is a measure of entanglement
and ${\cal D}$ is the set of disentangled states. In this approach it is
found that the entanglement measure for $N$-particle GHZ state is 
$E(\arrowvert \Psi(-) \rangle) = (N-1) \log 2$. For $4$-particle GHZ state
produced in the experiment the amount of `operator norm based entanglement'
is then $E(\rho) = 2.472 \log 2$.

Thus depending on the use of the quantum resource and `measures of 
entanglement' used, the measures of four-particle pseudo-pure 
GHZ state are different.
Though these measures does not exhaust all possible measures of entanglement
at least answers an important open questions related to
the {\em experiment} which is a leap forward in manipulating and 
understanding of multi-particle entanglement with trapped ions. 
This would help us to quantify the resource available at our disposal 
for future technology.

\vskip 2cm

\noindent
{\bf Note:} This paper was written around April 2000, soon after the
experimental production of $4$-particle GHZ state and finalized in Benasque
Science Center for Quantum Computation, June-2000, Spain. I have added only few
recent references. I thank G. Kar for bringing the reference [17] to 
my notice. But many recent references could be missing.



\begin{references}

\bibitem{mike} Nielsen, M. A. and Chuang, I. L. (2000), 
Quantum Computation and Quantum Information, {\it Cambridge University Press},
 Cambridge.

\bibitem{rb} R. Blatt, Nature {\bf 404}, 231-232 (2000).

\bibitem{ms} K. Molmer and A. Sorenson,  Phys. Rev. Lett.
{\bf 82}, 1835-1838 (1999).

\bibitem{sack} C. Sackett {\it et al}, Nature {\bf 404}, 256-259 (2000).


\bibitem{ved1} V. Vedral,M. B. Plenio, M. A. Rippin and
P. L. Knight, Phys. Rev. Lett. {\bf 78}, 2275-2279 (1997).

\bibitem{bprst} C. H. Bennett, S. Popescu, D. Rohrlich, J. A. Smolin, 
and A. V. Thapliyal, Phys. Rev. A {\bf 63}, 012307 (2001).

\bibitem{hhh} M. Horodecki, P. Horodecki, R. Horodecki, 
Phys. Rev. Lett. {\bf 84}, 2014 (2000).


\bibitem{pb} A. K. Pati and S. L. Braunstein (Unpublished notes 2000).

\bibitem{rfw} R. F. Werner, Phys. Rev. A {\bf 40}, 4277 (1989).

\bibitem{ls} M. Lewenstein and A. Sanpera, Phys. Rev. Lett.
 {\bf 80}, 2261-2264 (1998).


\bibitem{bet} C. H. Bennett, 
G. Brassard, S. Popescu, B. Schumacher, J. Smolin, and W. K. Wootters,
Phys. Rev. Lett. {\bf 76}, 722 (1996).

\bibitem{bdsw} C. H. Bennett, D. P. DiVincenzo, 
J. A. Smolin, and W. K. Wotters, Phys. Rev. A  {\bf 54}, 3824 (1996).


\bibitem{sam} S. L. Braunstein {\it et al},
Phys. Rev. Lett. {\bf 83}, 1054 (1999).

\bibitem{bp} S. L. Braunstein and A. K. Pati, Quant. Inf.
and Comp. {\bf 2}, 399 (2002).

\bibitem{em} B. G. Englert and N. Metwally, quant-ph/9912089 (1999).

\bibitem{meyer} D. A. Meyer and N. R. Wallach, quant-ph/0108104 (2001).

\bibitem{yuka}  V. I. Yukalov, Phys. Rev. Lett. {\bf 90} 167905-4 (2003).

\end{references}
\end{document}